\documentclass[runningheads]{llncs}
\usepackage[T1]{fontenc}
\usepackage{graphicx}
\usepackage[breaklinks]{hyperref}
\expandafter\def\expandafter\UrlBreaks\expandafter{\UrlBreaks
  \do\a\do\b\do\c\do\d\do\e\do\f\do\g\do\h\do\i\do\j%
  \do\k\do\l\do\m\do\n\do\o\do\p\do\q\do\r\do\s\do\t%
  \do\u\do\v\do\w\do\x\do\y\do\z\do\A\do\B\do\C\do\D%
  \do\E\do\F\do\G\do\H\do\I\do\J\do\K\do\L\do\M\do\N%
  \do\O\do\P\do\Q\do\R\do\S\do\T\do\U\do\V\do\W\do\X%
  \do\Y\do\Z}

\usepackage[nolist]{acronym}

\usepackage{booktabs}
\usepackage{tabularx}
\usepackage{amssymb} 
\usepackage{float}

\usepackage{tablefootnote}

\begin{document}
\title{Provenance of Adaptation in Scientific and Business Workflows -- Literature Review}
\titlerunning{Provenance of Adaptation -- Literature Review}
%
\author{Ludwig Stage\inst{1}, Julia Dahlberg\inst{2}, Dimka Karastoyanova\inst{1}}
%
%
\institute{Information Systems Group, University of Groningen, The Netherlands \email{\{l.stage,d.karastoyanova\}@rug.nl}\and
University of Groningen, The Netherlands\\ \email{j.i.k.dahlberg@student.rug.nl}}
\maketitle              
\begin{acronym}
	\acro{ACID}{Atomicity, Consistency, Isolation, Durability}
	\acro{BPEL}{Business Process Execution Language}
	\acro{BPM}{Business Process Management}
	\acro{BPMN}{Business Process Model and Notation}
	\acro{BPMS}{Business Process Management System}
	\acro{EPR}{Endpoint Reference}
	\acro{ESB}{Enterprise Service Bus}
	\acro{FAIR}{Findable Accessible Interoperable Reusable}
        \acro{MitM}{Machine-in-the-Middle}
        \acro{PAIS}{Process-Aware Information System}
	\acro{PoC}{Proof of Concept}
	\acro{SLA}{Service-level agreement}
	\acro{SOA}{Service-oriented Architecture}
	\acro{SOC}{Service-oriented Computing}
	\acro{RARE}{Robust Accountable Reproducible Explained}
   \acro{UDDI}{Universal Description, Discovery and Integration}
   \acro{UUID}{Universally Unique Identifier}
   \acro{WSDL}{Web Services Description Language}
   \acro{WfMS}{Workflow Management System}
   \acro{sWfMS}{Scientific Workflow Management System}

   \acro{DAG}{Directed Acyclic Graph}
   \acro{OPM}{Open Provenance Model}
   \acro{W3C}{World Wide Web Consortium}
   \acro{PROV-DM}{The PROV Data Model}
   \acro{RDF}{Resource Description Framework}
   \acro{PROV-O}{The PROV Ontology}
   \acro{OPMW}{Open Provenance Model for Workflows}
\end{acronym}

\setlength{\tabcolsep}{4pt}

\begin{abstract}

In the world of science new technology have opened up the possibility to rely on advanced computational methods and models to conduct and produce scientific research. An important aspect of scientific and business workflows is provenance - which refers to the information describing the production, history or lineage of an end product, which can also be data, digitalized processes and other not tangible artifacts. 
While there are already systems, tools and standards to capture provenance of data and workflows the provenance of adaptations/changes in workflows has not been addressed yet. 
In this paper we carry out a literature review to establish the state of the art on this topic and present our methodology and findings. Our findings confirm that provenance of adaptation has not been addressed adequately in the fields of business and scientific workflows. The two fields also have different motivation for recording the lineage of data or processes. While scientific workflows are interested in reproducibility and visualization, business workflows solutions are indirectly connected to compliance, exception handling and analysis. The adaptive nature of workflows in both fields is not reflected in the research on process provenance yet, as our results show. The use of standard provenance standards is also not wide spread.


\end{abstract}
\section{Introduction}
The technology is ever so quickly evolving in our society, which opens up new possibilities in many different areas but also pushes us to adapt to the new innovations. One of the areas affected by this change is science. New technology has opened up the possibility to rely on advanced computational methods and models to conduct and produce scientific research. With this new way of doing research new findings are possible, but new challenges are also presented. One of these challenges is presented in the aspect of reproducible research \cite{wilkinson_2016,mesirov_2010}. For in silico research to be deemed reproducible, there needs to be detailed information accessible about the software environment used, which data was used and produced and how every step was carried out. All of this information supports establishing the provenance of the research.

Provenance is not only crucial in scientific experiments but also in business workflows. In scientific and business workflows provenance describes the production or history of a data product. The capturing of provenance is a topic that has gained importance, at least since the increase of in silico research and simulated experiments with its handling of large amounts of data and additional levels of complexity. As a result of this there are different tools and systems to be used for capturing and managing provenance in scientific and business workflows.

There are many different aspects of provenance and one challenge is how to capture them all, or alternatively how to decide what aspects are needed to ensure reproducibility. One of these aspects that we will examine in this paper is the information of change or adaptation in a workflow during runtime. This change may occur for example when a scientist decides to change the software used for a part of the workflow, or the dataset used receives some additional data-points. These workflow changes are crucial for the trial-and-error manner of conducting their experiments and their research, making it important to capture this information. As of today, many of these important details of the process might get lost, which in turn would affect the reproducibility aspect of the experiment.\\

In this paper we report the findings made in a systematic literature review on this very topic. We will examine different methods and tools used to capture and visualize provenance, as well as what is missing in this area. With this information and potential future solutions we will identify future challenges, research topics, and open questions for investigation.

\section{Background Information}
\subsection{Provenance}
In scientific research all aspects of the research process, including methodologies, data collection procedures, data and workflows used are expected to be established and captured in such a way that the reproducibility of the research results can be ensured. Reproducibility is essential in order to replicate the research and verify its findings, thereby promoting trust in scientific inquiry.

The collection (also referred to as capture) and processing of provenance are important in various settings, e.g., to assess quality, to ensure reproducibility, or to reinforce trust in the end product.

This has arisen as one of the challenges of eScience, as reproducing an advanced computational experiment needs a lot of specific data and methods which can be difficult to document and keep track of during the research process. This is where the concept of provenance information becomes a crucial aspect \cite{wilkinson_2016,mesirov_2010}. 

\subsubsection{Provenance Taxonomy}

In this report, we base our definition of provenance on the work by Herschel et al. \cite{id:20_herschel_2017}, according to which "\textit{provenance} refers to any information that describes the production process of an end product, which can be anything from a piece of data to a physical object". As provenance information encompasses meta-data about entities,
data, processes, activities, and persons involved in the production process, provenance information can be seen as meta-data describes a production process rather than the describing data \cite{id:20_herschel_2017}.

This definition serves as a foundation for exploring the different features of provenance information. One of these features is the granularity, which indicates the amount of detail in the provenance. There are two different granularities identified according to \cite{id:20_herschel_2017}: fine-grained and coarse-grained granularity.

Another feature is the provenance-type. There are four different types of provenance depending on the area of use: provenance meta-data, information systems provenance, workflow provenance and data provenance \cite{id:20_herschel_2017}. In this paper we are going to focus on provenance of scientific- and business workflows. 

The last feature of provenance, and the one we will examine the most in this report, is the provenance-form. In \cite{id:20_herschel_2017} there were three different provenance-forms identified: prospective, retrospective and evolution provenance. 

Prospective is the provenance-information that describes the structure or static context of a workflow, which means that it is not dependent on the input or execution of the workflow. Prospective provenance is captured by process or workflow models from the BPM domain. 

Retrospective provenance on the other hand contains the specific execution information of a workflow. This includes information of the execution of every workflow step and the environment, as well as the accessed or produced resources. Retrospective provenance of workflows in BPM is captured in process logs that are typically captured by the process execution environment and record the historical information about process instance executions. 

The last form, evolution provenance describes the changes between two different versions of a workflow. In BPM a similar approach is used to identify evolutionary changes in process models, which is a modeling phase concern. Notably, the current taxonomy of workflow provenance does not consider workflow changes that happen during the execution phase on scientific or business workflows.

To close this gap, in this report we are going to examine the state-of-the-art of provenance of adaptation. This term is used by \cite{id:17_2023} to describe the adaptation or change in a workflow, more specifically ad-hoc changes and changes during process runtime, which in turn will enable the provenance of adaptive workflows. 

\begin{figure}[htb!]
    \centering
    \includegraphics[width=\textwidth]{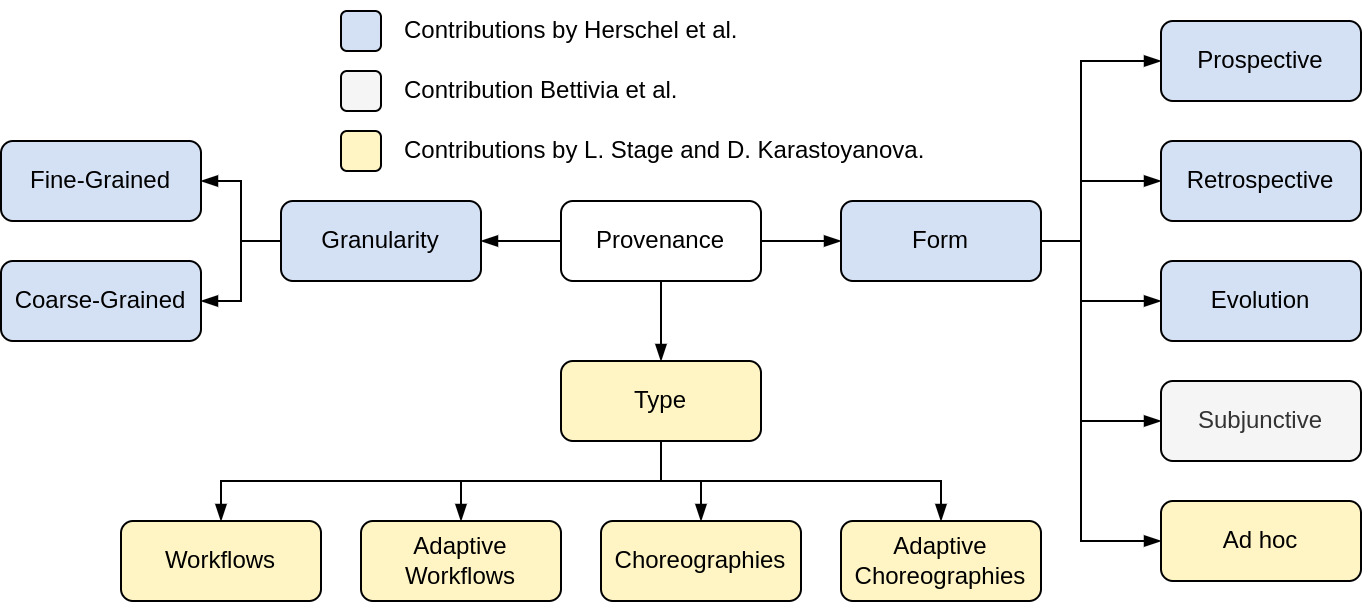} 
    \caption{Workflow Provenance types taxonomy \cite{id:20_herschel_2017,id:17_2023,id:6_LACK_2023}}
    \label{fig:tax}
\end{figure}

To account for provenance of change therefore, the forms of provenance originally mentioned in \cite{id:20_herschel_2017} has been extended in \cite{id:17_2023} with ad-hoc provenance
and we extent it here to include subjunctive provenance \cite{id:6_LACK_2023}
as depicted in \autoref{fig:tax}.
Furthermore, the workflow provenance types have been added to the taxonomy and subdivided into four groups to distinguish between workflows and choreographies and adaptive workflows and choreographies.


\subsubsection{Provenance Capturing and Visualization}
Depending on the use case, different data sources, processing pipelines and parties involved, the capturing and managing of provenance can look very different. As there are also currently no adopted global standards for provenance, the specific data collected by a provenance capture tool can vary greatly. The capture can be done either by the \ac{WfMS}, a separate tool connected to the \ac{WfMS}, or a standalone tool. The specific data actually captured by either method also varies. Some tools may only collect partial provenance, the types and forms of provenance can vary, or if the data collected is fine-grained or coarse-grained.

The visual representation of the provenance can also vary between different approaches, as there is not a given standard. Visualization can be anything from a simple log in a document to complex and in-depth graphs. 

We will present our findings about capturing and visualization later in this study.

\subsection{Business and Scientific workflows}

Business workflows are defined by \cite{business_workflow_1995} as a description of tasks, only specifically within a business process. These tasks are described on a theoretical level to enable the understanding of the business process, as well as for evaluating, and redesigning the process. The business workflow may also include information describing the process requirements \cite{Leymann2000_ProductionWorkflow,Weske19}.

A scientific workflow shares many similarities with a business process, and can even be modeled as one. According to \cite{scientific_workflow_2009} the scientific workflow is defined as follows: a description of a scientific process, breaking it down into a series of smaller steps. It is constructed of components called "tasks" and "dependencies". The tasks represent computational steps used in for example data analysis and simulations, and are usually arranged and executed in a specific order. The dependencies represent the relationships between these tasks in the workflow. For example the output of one task might be required as the input of another task, so tasks have the possibility of being dependent on each other.

\section{Methodology}
The goal with this paper is to examine the current state of provenance in scientific and business workflows, and more specifically provenance of adaption. For this purpose we have formulated a selection of research questions. To explore this topic and answer the research questions the chosen methodology is the systematic literature review approach. The review approach will follow the guidelines of \cite{review_guide_2007}.

\subsection{Research question}
To explore the topic of provenance of adaptation in scientific and business workflows and how to capture it we have identified one main research question (RQ1) with five sub-questions (cf. \autoref{tab:research-questions}).

\begin{table}[htp]
\centering
		\caption{Research questions: Main research question RQ1, sub-questions RQ2-RQ6}
		\label{tab:research-questions}
\begin{tabularx}{\textwidth}{c X}
	\toprule
        \textbf{RQ1} & How do we capture provenance information of adaptive scientific and business workflows?\\
        \textbf{RQ2} & What information should be captured, and what is currently being captured?\\
        \textbf{RQ3} & How is such information being used?\\
        \textbf{RQ4} & How is such information being visualized?\\
        \textbf{RQ5} & What are the available works and tools for this purpose?\\
        \textbf{RQ6} & What is currently missing within this subject?\\
	\bottomrule
\end{tabularx}
\end{table}

\subsection{Scope}
The topic was defined as provenance of adaptation in regards to scientific and business workflows. Specifically, we will examine the workflow-provenance type and explore all forms and levels of granularity of provenance that we encounter. We will include papers on provenance in scientific- and business workflows, encompassing both those that involve adaptations and those that do not, as these can be used to answer the sub-questions and still provide valuable insight to the main research question. We will include papers related to scientific and business workflows - regardless of the specific field of study. We will also include papers on business processes and scientific choreographies shall we encounter them. 

Our goal is to investigate the current state of the art of this topic, using \cite{id:20_herschel_2017} as a foundation. Therefore, we aim to use papers published subsequent to the release of this report in this review, or potentially older studies they may have overlooked or excluded for any reason. 

\subsection{Literature search}
This is the step were we define the search terms and perform our search on each of the selected search sources. The goal is to find and collect every paper that is related to our topic and can be used to answer the research-questions (cf. \autoref{tab:research-questions}). The search terms and search sources used are listed in \autoref{tab:search-terms} and \autoref{tab:search-sources}. The iterations of the literature search and the selection process is documented in \autoref{tab:iterations}. The final step involves the analysis of references. During this stage, the references from each paper are reviewed to identify additional relevant studies to include in the review, resulting in an increase in the total number of papers.

\begin{table}[H]
    \centering
    \caption{Search terms for provenance in workflows}
    \label{tab:search-terms}
    \begin{tabularx}{\textwidth}{c X}
        \toprule
        \textbf{Query} & \textbf{Description} \\
        \midrule
        1. & (provenance AND (adaptive OR adaptation) AND ((scientific OR business) AND workflow) \\
        2. & provenance AND (scientific workflows OR business workflows OR adaptive workflows) \\
        3. & provenance AND (adaptive workflows OR flexible workflows) \\
        \bottomrule
    \end{tabularx}
\end{table}

\begin{table}[htp]
    \centering
    \caption{Search sources for literature review}
    \label{tab:search-sources}
    \begin{tabularx}{.75\linewidth}{c X}
        \toprule
        \textbf{Source} & \textbf{Description} \\
        \midrule
        1 & Web of Science (WoS) \cite{web_of_science}\\
        2 & Scopus \cite{scopus}\\
        3 & IEEE Xplore \cite{ieee} \\
        4 & Digital Bibliography \& Library Project (DBLP) \cite{dblp} \\
        \bottomrule
    \end{tabularx}
\end{table}
\begin{table}[htp]
    \centering
    \caption{Literature Search Process:
    Number of papers before and after each step (Detailed steps: \url{https://drive.google.com/file/d/1hPSRTp8wmwdQRqIiNP1Z0lCT-yhleKA0/view})}
    \label{tab:iterations}
    \begin{tabularx}{\textwidth}{l l l}
        \toprule
        \textbf{Step} & \textbf{Before} & \textbf{After}\\
        \midrule
        Papers per source & 0 & WoS: 
307, Scopus: 503, IEEE: 233, DBLP: 122\\
        Initial search & 1165 & 68\\
        Review process & 68 & 16\\
        Analyzing references & 16 & \textbf{22}\\
        \bottomrule
    \end{tabularx}
\end{table}

\section{Result}

The selected studies for the review consisted mainly of papers focused on scientific workflows, as only 5 out of the total 22 reviewed papers covered business workflows or processes. However as both workflows are similar in structures, most information are applicable in both fields. This also means that the same tools and methods can generally be used for both scientific- and business workflows. Because of this there will not be a separation of these studies based on their applied field.

The studies were of many different types, so to get a better understanding of how they differed and which insight we can gather from them we are going to divide them into different categories. Note that some of these studies can belong to more than one category.

The first category, which is also the most common type, is the \emph{Conceptual studies}. These studies build upon previous research, using it as a foundation to propose new concepts or ideas. These concepts have not actually been fully implemented or tested yet, at the time of this study's publication.

The next category is \emph{Empirical studies}. This study type analyses an already implemented tool or method that has been tested and analyzed.

As there was a limited amount of papers related to provenance of adaptation, this was not selected as a requirement and other studies not directly related to adaptation were also included in the review. As such, we introduce a category called \emph{Adaptation-related studies}, to indicate studies containing discussion about provenance of adaptations or adaptive workflows.

The last category is called \emph{General studies}, and this includes studies on the general topic of provenance or workflows, as well as review papers and surveys.
The contributing studies for each category can be seen in \autoref{study-types}.\\

\begin{table}
\centering
\caption{Study types}\label{study-types}
\begin{tabularx}{.7\textwidth}{l X}
\toprule
\textbf{Type} & \textbf{References}\\
\midrule
Conceptual & \cite{id:2_2018,id:5_PRISM_2023,id:6_LACK_2023,id:8_stage_2024,id:10_2017,id:17_2023,id:18_2019,id:1_2014,id:7_samba_2020,id:9_2021,id:13_2011,id:14_2010,id:16_2022,id:19_avocado_2016} \\
Empirical & \cite{id:12_2006,id:15_2008,id:22_2021}\\
Adaptation-related & \cite{id:2_2018,id:5_PRISM_2023,id:6_LACK_2023,id:8_stage_2024,id:10_2017,id:17_2023,id:18_2019,id:20_herschel_2017}\\
General studies & \cite{id:3_2017,id:4_2008,id:11_2023,id:20_herschel_2017,id:21_2016}\\
\bottomrule
\end{tabularx}
\end{table}

Within the studies in which provenance of adaptation were discussed, the adaptation aspect was sometimes the main focus of the paper but in others it may only have been discussed briefly. There was also different terms used to describe provenance of adaptation, and the focus of adaptation would also differ. The previously mentioned form of evolution provenance was used in \cite{id:20_herschel_2017}. The term "provenance of change" was used in \cite{id:8_stage_2024,id:17_2023} and "provenance of adaptation" was used in \cite{id:17_2023}, both used to describe provenance of adaptations or changes in workflows. In \cite{id:6_LACK_2023} the term "subjunctive provenance" was used to describe not the adaptation itself, but other possible changes or events that could have happened but did not occur. In the papers \cite{id:2_2018,id:5_PRISM_2023,id:10_2017,id:18_2019} there were no specific term used to describe provenance of adaptation.\\

The studies not directly related to adaptation were also included, if they fulfilled the selection criteria and the quality assessment. These papers were used to gain further insight into the topic of provenance in scientific and business workflows, as well as to answer the respective research questions. The sub-questions in which these papers were relevant are: RQ2, RQ3, RQ4 and RQ5 (cf. \autoref{tab:research-questions}). As such, these papers will mainly provide insight into which provenance information is currently being captured, how it is being used and visualized, and what tools are being used for this purpose.

\subsection{Capturing Provenance Information}

When it comes to the capturing of provenance, in answer to RQ2 (cf. \autoref{tab:research-questions}), there are two different types of studies encountered during the review that are deemed most relevant: \emph{Conceptual studies} and \emph{Empirical studies}. The \emph{General studies} can also be used to gain insight into the general state of provenance capturing. As such, for this section there is a division of these categories to describe the capturing of provenance.\\
%
%
\subsubsection{Information currently being captured}

The information retrieved from the studies on this topic 
vary depending on the context and the specific workflow in question. However, we have gathered from the reviewed papers that there are specific provenance types and forms that are more commonly collected.

In \cite{id:20_herschel_2017}, workflow provenance is defined as meta-data collected for a workflow process that can be derived from the input, output, the model and the parameters of the workflow process. If we look more into detail at the workflow provenance captured we observe that both granularities are reported to be captured, depending on the use cases of the provenance. There are also mainly two forms that are mentioned to currently being captured, namely prospective and retrospective provenance. The only mentions of evolution provenance being captured has been in \cite{id:20_herschel_2017} which states that Kepler \cite{kepler} and VisTrails \cite{vistrailswiki} provides support for evolution provenance. 

\subsubsection{Information to be captured}

To answer the question of what information should be captured (to enable the capture of adaptation), we have to look into the papers of the category \emph{Adaptation-related studies}. It is inevitable that provenance is needed, but we will look into what exactly differs between provenance and the provenance of adaptation. In \cite{id:2_2018} it is mentioned that capturing the provenance of the adaptation steps made within the workflow is necessary, and to further explore this we will examine exactly what information this entails. \\

Some studies identify that there are new forms of provenance needed to be collected to enable adaptation, additionally to the forms defined by \cite{id:20_herschel_2017}. As mentioned before there are three provenance forms identified: prospective, retrospective and evolution provenance. In \cite{id:6_LACK_2023} a fourth form of provenance is also identified, namely subjunctive provenance. This form of provenance describes what could happen during the implementation of a process. The authors provide an effective demonstration of the various provenance forms using the process of building an IKEA table as an example.

Using this example, subjunctive provenance could be explained by the fact that some tool described in the IKEA table manual is not included in the kit, which may lead to the customer using another tool or method to assemble the different parts of the table. It could also be used the other way around, if such a modification has taken place subjunctive provenance may be used to describe what could have happened would this change not have occurred. Subjunctive provenance is thereby needed to identify potential branches that could emerge within a workflow, or to look back at a process and see what could have happened if other choices were made.\\
 
In \cite{id:17_2023} there has also been an additional form of provenance identified, namely ad-hoc provenance. Provenance of change has been subdivided in \cite{id:17_2023} into the forms evolution provenance (originally mentioned in \cite{id:20_herschel_2017}) and ad-hoc provenance. The difference between these two are identified as evolution provenance describe adaptations to the workflow model itself, and ad-hoc provenance describes the adaptations in a workflow instance. Using the same example as in \cite{id:6_LACK_2023} with the IKEA table, evolution provenance can be compared to the company replacing the provided screws in the table package for nails. This would enable customers to use a hammer for assembly instead of a screwdriver, generating a change in the building process for the table, affecting all customers. Ad-hoc provenance on the other hand could represent one specific customer not finding the right sized screwdriver, and therefore opting to exchange the screws for nails that they already possessed. As such, this does not affect the general building process for the table, only for this specific instance of it.\\

\subsection{Usage and Visualization}

There are many use cases for provenance mentioned in the studies, but the most commonly discussed utilization is for the purpose of reproducibility within workflows. Within the scientific community reproducible research is fundamental and within scientific and business workflows provenance information is the very basis of this important feature. 

Within the area of collaborative adaptive workflows, provenance of adaptation is especially required to ensure reproducibility according to \cite{id:18_2019,id:17_2023}. In these papers the authors assert that in order to reproduce data processing pipelines between collaborating organizations it is necessary to record a description of the change being performed, as well as the new and the old version of the workflow.\\ 

\begin{figure}[hbt!]
    \centering
    \includegraphics[width=\textwidth]{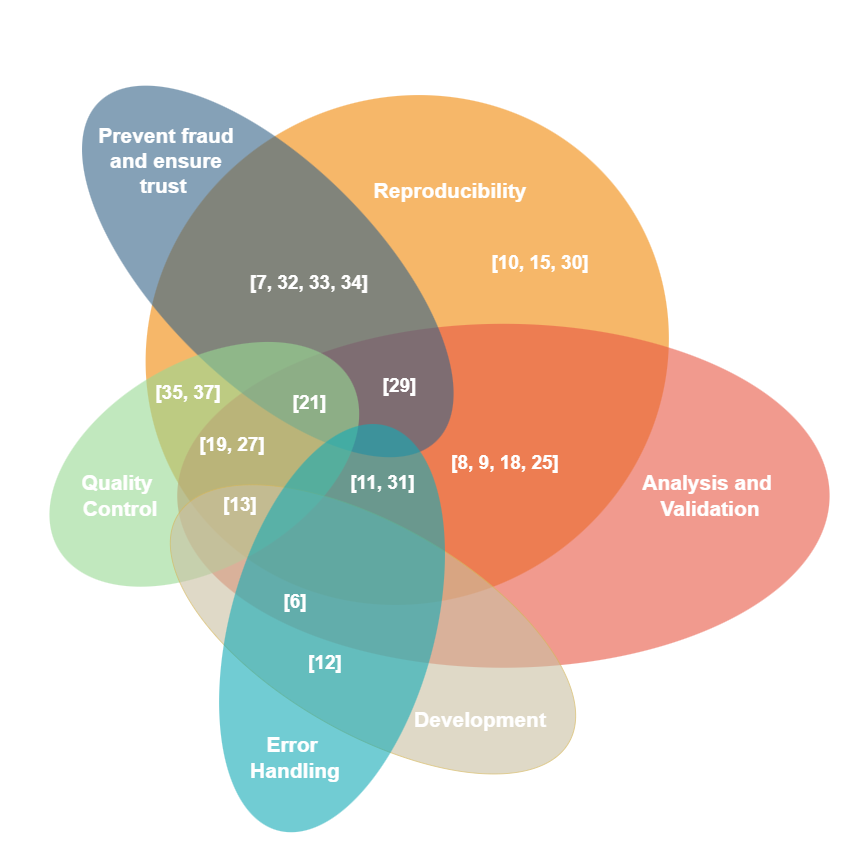} 
    \caption{Use cases in the literature}
    \label{fig:use_cases}
\end{figure}

Except for reproducibility there are multiple other use cases for provenance. \autoref{fig:use_cases} presents the use cases discussed in the studies, along with the corresponding studies where these use cases were mentioned. Many use cases stated in \autoref{fig:use_cases} are related to each other. For example: analyzing and validating the data and its derivation can help researchers explain unexpected results or identifying and handling errors. This makes it possible to further prevent fraud and ensure trust within the collaborators, as all the results are traceable. Moreover, handling errors and preventing fraud helps maintain quality throughout the experiments thereby facilitating quality control. As such, collecting and documenting provenance enables a lot of use cases with different advantages for researchers.\\

When it comes to visualization of provenance there are different approaches. The majority of studies discussing visualization mention some form of graph-representation to visualize the provenance. The use of a \ac{DAG} is described in \cite{id:16_2022,id:15_2008,id:19_avocado_2016,id:12_2006}, although these graphs are constructed in different ways. For example the visualization graphs in \cite{id:15_2008} are made of graphs where each node represents data, collection and parameter items produced by (or provided to) the workflow run. The edges maps each node to the set of nodes and events involved in its creation, showing how one item is derived from others or influenced by specific events. 

The graph described in \cite{id:12_2006} are described as a rooted tree, where each node represents a version of the workflow. The edges between these nodes then describe the action needed for the derivation of one version from another. Some tools and methods discussed in the studies provide support for PROV, which is a provenance standard that can implicitly be visualised as a graph. 

Some less common visualization techniques mentioned are the use of provenance reports \cite{id:7_samba_2020}, and a database-visualization \cite{id:15_2008}.

\subsection{Tools and works}

There has been many different tools and methods encountered during the review, and once again it is necessary to divide them into two categories: implemented tools and methods, and conceptual tools and methods. The implemented tools are tools and methods that are fully implemented and officially released. Conceptual tools are defined as those that are either not fully implemented or consist of elements that are still only theoretical. 

In the following we look at the implemented tools and methods as these will give us a general view of what is available at the moment of writing. We  will also present conceptual tools and methods available that are related to capturing the provenance of adaptations. The tools discussed in this section may be discussed in great detail in the studies or simply be briefly mentioned. If briefly mentioned, additional research has been made to answer the relevant questions such as: What is the main purpose of this tool? Which provenance standards are supported? Etc.

\subsubsection{Implemented tools and methods}

During the review, various provenance tools designed for different purposes were encountered. A collection of the reviewed implemented tools and methods is illustrated in \autoref{imp-tools}. The tools included in these tables are the tools mentioned in the studies that contain some support for provenance. It is stated what the main purpose of the tool is in regards to provenance, and whether there is support for some provenance standard. Under "Type," it is specified whether the tool is a \ac{WfMS}, incorporated with a \ac{WfMS}, or a standalone tool. "Last Change" indicates the most recent year when the code in the open repository was updated. Lastly it is stated in which papers this tool or method was mentioned.

\renewcommand{\arraystretch}{1.2}
\begin{table}
\centering
\caption{Implemented tools.
C: Capture, M: Manage, V: Visualize, E: Edit}\label{imp-tools}
\makebox[1 \textwidth][c]{
\resizebox{1.3 \textwidth}{!}{
\begin{tabular}{l c c c c c l }
\toprule
\textbf{Name} & \textbf{Purpose} & \textbf{Standard} & \textbf{Type} &  \textbf{Last Change} & \textbf{References}\\
\midrule
Kepler & C & PROV, OPM & \ac{WfMS} &  2021 & \cite{id:17_2023,id:16_2022,id:15_2008,id:9_2021,id:4_2008,id:3_2017,id:20_herschel_2017}\\

Prov Viewer\footnote[1]{} & V & PROV & Standalone &  2020 &  \cite{id:17_2023,id:22_2021}\\

ProvViz\tablefootnote{If the standard is fully implemented and did not change since the last change of the tool, no change is ok.} & V, E & PROV, RDF & Standalone & 2021 & \cite{id:17_2023,id:22_2021}\\

ReproZip & C, M & None & Standalone & 2024 & \cite{id:10_2017}\\

Taverna & C, M & PROV, OPM & \ac{WfMS} &  2020 & \cite{id:9_2021,id:4_2008,id:3_2017,id:18_2019,id:20_herschel_2017} \\

VisTrails &  C, M, V & OPM & \ac{WfMS} & 2017 & \cite{id:10_2017,id:19_avocado_2016,id:12_2006,id:9_2021,id:4_2008,id:20_herschel_2017,id:21_2016}\\

\bottomrule
\end{tabular}
}
}
\end{table}

As can be seen in \autoref{imp-tools}, most of these tools have been developed for the main purpose of managing provenance, collection of provenance or for visualization. The most mentioned tools are: Kepler \cite{kepler}, VisTrails \cite{vistrailswiki} and Taverna \cite{tavernaprovenance}. Of these tools, only Kepler is still being maintained as both VisTrails and Taverna are no longer maintained according to their respective websites or specifications. According to \cite{id:20_herschel_2017}, evolution provenance is supported by Vistrails and Kepler, but none of the tools are reported to support ad-hoc provenance. Moreover, all the implemented tools mentioned in this table were found to be open source.

\subsubsection{Conceptual tools and methods}

As previously stated, many of the studies used in this review are conceptual studies, introducing a conceptual or theoretical idea that has not yet been fully implemented. These concepts are important to distinguish from the implemented tools and methods, as these have not yet been thoroughly tested. These concepts do however contain potential and may provide methods that could enable the capture and management of provenance of adaptation, which is something that the already implemented tools and methods have not been able to fully satisfy. Therefore this category is very important to be able to answer the main research question. 

An overview of the conceptual tools and methods encountered during this review are presented in \autoref{conc-tools}. 

The columns are the same as \autoref{imp-tools}, with the addition of the "Adaptation" and "Open Source" columns, indicating whether or not there is expressed support for adaptations, and if the tool is open source. Instead of noting the last year of change, we also consider the "Last Work". Since not all of these tools are implemented, this can also indicate when the most recent paper discussing the tool was published.

\renewcommand{\arraystretch}{1.2}
\begin{table}
\centering
\caption{Conceptual tools. C: Capture, M: Manage, V: Visualize, S: Storage}\label{conc-tools}
\makebox[1 \textwidth][c]{
\resizebox{1.3 \textwidth}{!}{
\begin{tabular}{ l  c  c  c   c   c   c   c }
\toprule
\textbf{Name} & \textbf{Purpose} & \textbf{Adaptation} & \textbf{Standard} & \textbf{Type} & \textbf{Open source} & \textbf{Last Work} & \textbf{References}\\
\midrule
AVOCADO & V & \texttimes& None & Standalone & \texttimes& 2017 & \cite{id:19_avocado_2016,id:22_2021}\\

PRISM  & C, M, V, S & \checkmark & None & Standalone & \checkmark & 2023 & \cite{id:5_PRISM_2023}\\

Provenance Holder & C, M, (V), S & \checkmark & PROV & Incorporated & \checkmark & 2024 & \cite{id:18_2019,id:17_2023,id:8_stage_2024}\\

ProvSearch  & M, S & \texttimes& None & Incorporated & \texttimes& 2014 & \cite{id:1_2014}\\

SAMbA-RaP  & C, S & \texttimes& PROV & Incorporated & \checkmark & 2020 & \cite{id:7_samba_2020}\\

Secure scientific workflow data provenance framework & C, S & \texttimes& ProVOC, RDF & Standalone & \texttimes& 2023 & \cite{id:11_2023}\\
\bottomrule
\end{tabular}
}
}
\end{table}

From \autoref{conc-tools} we can derive that only two tools currently indicate some form of support for provenance of adaptation: PRISM and Provenance Holder. PRISM introduces a design enabling the storing of provenance on a decentralized ledger in the form of a blockchain. It is meant to provide a flexible framework to support the dynamic nature of scientific workflows. To enable the support for workflow modifications, PRISM uses Invalidation and Modification Blocks. A scientist needs to get the agreement of at least 50\% of the scientist to be able to submit an invalidation transaction, or a modification transaction. After this a Data Invalidation or Modification Block will be added to the ledger. When the Invalidation Block gets propagated into the network, the relevant data will be flagged as invalid. Following this invalidation, the workflow task which produced the invalid data will be recomputed. As a consequence of this, every other workflow task relying on the invalid data must also be invalidated and recomputed. If there is a need for addition or subtraction of workflow tasks, a Modification Block is to be used. This provides scientists the possibility to adapt workflows during the experiments. The use of blockchain in this implementation simultaneously ensures comprehensive and transparent provenance records \cite{id:5_PRISM_2023}, however the provenance of workflow adaptations is a built in feature for this specific tool and does not allow for integration with other WfMS or process-aware information systems. The tool's main goal has been to enable coordinated adaptations, which in addition can be recorded as part of the provenance information.

According to \cite{id:17_2023} the Provenance Holder will have support for the coarse-grained granularity, and the prospective and evolution provenance forms. More specifically it will support both workflow evolution provenance and provenance of ad-hoc workflow change, according to the previously introduced definitions. It is also stated that support for retrospective and fine-grained provenance could also be enabled with detailed workflow execution traces. The Provenance Holder collects provenance on a very detailed level during the execution and adaptation of workflows. The data collected is highly specific and includes information about each individual activity within the workflows. The specific provenance information for a workflow adaptation might consist of a description of the specific modification, the new version of the workflow, as well as a reference to the old (preceding) version. The adaptations to the workflow model are stored in an object type called provenance information object for adaptation. When an instance migration occurs, the entire workflow model is captured. However, if ad-hoc workflow changes happen, only the specific modification is captured. Before storing the provenance the data gets validated. Similarly to PRISM the Provenance Holder suggests the use of an immutable public ledger, for example blockchain, but instead of storing the provenance it is used for time-stamping. The primary use for this time-stamping is to ensure for example that a fact was known or an action happened at a certain time.

In \cite{id:10_2017} they also describe a method of descriptor-space containing all the necessary parameters for the workflow within descriptors. For each descriptor there is a name, a value and a decay-parameter. This decay-parameter is tracking whether the parameters are accessible, as well as if there are changes within the parameters. For the workflow to be fully documented and reproducible all parameters need to be known or stored. Therefore this descriptor-space method makes it possible to analyse whether or not a workflow is fully reproducible and if there has been any changes made to the parameters. One example of this could be if a workflow is depending on data subtracted from a database that is continuously changing, the decay-parameter will indicate that a change has been made and this adaptation could be documented if the system is extended further in this direction. 

The work \cite{id:11_2023} presents a scientific workflow data provenance framework based on the provenance model ProVOC and blockchain. Using blockchain, the framework ensures that the provenance information has not been altered or tampered with. 

\subsubsection{Provenance standards}

There are many different provenance representation models or standards used to represent provenance. Initially, \ac{RDF} \cite{rdf} was used to present provenance. \ac{RDF} is a data model developed by \ac{W3C} for meta-data and a key element of the semantic web. \ac{RDF} represents data as triples, each consisting of a subject, predicate, and object. 

Following this, one of the first models developed specifically for provenance in 2007 was the \ac{OPM} \cite{opm}, which is a Graph-Based Model. \ac{OPM} did however come with some problems and as a result another more refined and detailed standard was proposed by \ac{W3C}, called PROV \cite{w3c}. The very core of the PROV standard is the data model \ac{PROV-DM}. \ac{PROV-DM} can also be mapped to \ac{RDF} via PROV-O. 

Another notable model is \ac{OPMW}, which serves as an extension of the \ac{OPM} standard specifically designed to describe the provenance of scientific workflows. Also highlighted in the studies is ProvOne \cite{provone}, which also represents scientific workflow provenance. Notably, ProvOne is compatible with \ac{PROV-DM}, ensuring interoperability and flexibility in how provenance information is captured and utilized across different systems. Together, these models provide robust frameworks for managing the complexities of scientific workflows.

According to \cite{id:11_2023} \ac{PROV-DM}, \ac{OPMW} and ProvOne all have the ability to capture, store, and search the provenance of a workflow, as well as trace it in a standard, machine-readable format. ProvOne is also reported to support three forms of provenance: prospective, retrospective and evolution \cite{id:20_herschel_2017}.

\section{Discussion}

In this paper we have examined the state of the art of provenance in (adaptive) scientific and business workflows and answered the stated research questions on the topic. We have discussed two different forms of adaptive provenance: evolution and ad-hoc. To enable the capture of provenance information of adaptive scientific- and business workflows, it is important that we provide support for the capture of both evolution and ad-hoc provenance. As the most common use-case of provenance is reproducibility, it is crucial to enable the capture and managing of all forms of provenance, including provenance of adaptation. Without this we can not ensure full reproducibility for adaptive workflows. 

Some implemented tools (Kepler and Vistrails) do provide support for the capture of evolution provenance, but no support of ad-hoc provenance has been reported. There are also tools under development with reported support for provenance of adaptations, such as PRISM \cite{id:5_PRISM_2023} and the Provenance Holder \cite{id:17_2023}, although they are not fully implemented yet. The Provenance Holder reports intended support of both evolution and ad-hoc provenance, while PRISM does not define the supported provenance forms and focuses on capturing provenance information including collecting information about adaptations; the approach has however no specific focus on provenance of adaptation but rather on enabling adaptation and capturing immutable records of adaptations that have been carried out.

In the BPM field the term provenance is not typically used and the motivation for keeping the historical information regarding process executions is for the purposes of process analysis, discovery, compliance and improvement. There are approaches that can be used to capture or identify changes in execution logs that are based on either process mining or other ways to analyze process logs, however there is not comprehensive method to ensure provenance of adaptive workflows. For more on that topic, the reader is referred to the literature review by \cite{Turgay2022}.

With respect to the visualization of provenance of adaptations, there are tools available that supports the PROV standard (ProvViz \cite{id:22_2021} and Prov Viewer \cite{provviewer_2018}). Since this standard is flexible enough to capture the provenance and the adaptations, these tools are already compatible with provenance of adaptation, provided it is represented in a PROV format.

\subsection{Threats to validity}
One of the primary threats to the validity of this literature review is the relatively small number of papers available on the topic of provenance of adaptation. This limitation can impact the comprehensiveness of our findings as well as potential biases in the results. As provenance of adaptation is a relatively new and evolving area of research, the literature is still developing and some studies might be in progress or unpublished. Furthermore, due to the different focus and motivation of research in the two fields of scientific and business workflows, methods and techniques suitable for enabling provenance of adaptation may not use terminology common the discussions around provenance, hence our search string may not have been broad enough to consider such works.

\section{Conclusions}




While there are different use cases for provenance, our findings are predominantly located in the category of reproducibility within workflows.

We could identify different dimensions to provenance -- granularity (coarse- and fine-grained), form (Prospective, Retrospective, Evolution, Subjunctive, and ad hoc), type (Workflows, Adaptive Workflows, Choreographies and Adaptive Choreographies) -- and that they can be captured by different tools, implemented and conceptual. However, there is a gap between the information currently being captured and the information needed to support all dimensions, especially for provenance of adaptation and ad hoc provenance.

As more forms of provenance are identified, the volume of provenance information that needs to be captured, stored, and visualized also increases. This drives the development of new tools and methods capable of managing these complex data systems. Another key area for future work is determining which specific provenance information is necessary to ensure reproducibility or to meet the particular needs of the intended use case.
At the moment there is no tool which already supports for the capture, management, storage and visualization of provenance of adaptation and especially ad hoc provenance and is ready to use, yet. Only one line of research is to be projected to fill this gap, so far.
Further the relation between evolution, ad hoc, and subjunctive provenance are also a promising direction for future research.

Overall the need of an integrated method (and its supporting implementation) for capturing, managing, storing and visualizing provenance of adaptation and especially ad hoc provenance is clearly recognizable.

\bibliographystyle{splncs04}
\bibliography{ref}
\end{document}